# Reconfigurable Multifunctional van der Waals Ferroelectric Devices and Logic Circuits

Ankita Ram[1], Krishna Maity[1], Cédric Marchand[2], Aymen Mahmoudi[3], Aseem Rajan Kshirsagar[4], Mohamed Soliman[1], Takashi Taniguchi[5], Kenji Watanabe[6], Bernard Doudin[1,7], Abdelkarim Ouerghi[3], Sven Reichardt[4], Ian O'Connor[2]* and Jean-Francois Dayen[1,7]*.

1. Université de Strasbourg, IPCMS-CNRS UMR 7504, 23 Rue du Loess, 67034 Strasbourg, France.
2. École Centrale de Lyon, 36 Avenue Guy de Collongue, Ecully 69134, France.
3. Université Paris-Saclay, CNRS, Centre de Nanosciences et de Nanotechnologies, 91120, Palaiseau, France.
4. Department of Physics and Materials Science, University of Luxembourg, Luxembourg 1511, Luxembourg.
5. Research Center for Materials Nanoarchitectonics, National Institute for Materials Science, 1-1 Namiki, Tsukuba 305-0044, Japan
6. Research Center for Electronic and Optical Materials, National Institute for Materials Science, 1-1 Namiki, Tsukuba 305-0044, Japan
7. Institut Universitaire de France, 1 rue Descartes, 75231 Paris cedex 05, France

ABSTRACT

Emerging reconfigurable devices are fast gaining in popularity in the search for next-generation computing hardware, while ferroelectric engineering of the doping state in semiconductor materials has the potential to offer alternatives to the traditional von-Neumann architecture. In this work, we combine these concepts and demonstrate the suitability of Reconfigurable Ferroelectric Field-Effect-Transistors (Re-FeFET) for designing non-volatile reconfigurable logic-in-memory circuits with multifunctional capabilities. Modulation of the energy landscape within a homojunction of a 2D tungsten diselenide ($WSe_2$) layer is achieved by independently controlling two split-gate electrodes made of a ferroelectric 2D copper indium thiophosphate ($CuInP_2S_6$) layer. Controlling the state encoded in the Program Gate enables switching between p, n and ambipolar FeFET operating modes. The transistors exhibit on-off ratios exceeding $10^6$ and hysteresis windows of up to 10 V width. The homojunction can change from ohmic-like to diode behavior, with a large rectification ratio of $10^4$. When programmed in the diode mode, the large built-in p-n junction electric field enables efficient separation of photogenerated carriers, making the device attractive for energy harvesting applications. The implementation of the Re-FeFET for reconfigurable logic functions shows how a circuit can be reconfigured to emulate either polymorphic ferroelectric NAND/AND logic-in-memory or electronic XNOR logic with long retention time exceeding 10^4 seconds. We also illustrate how a circuit design made of just two Re-FeFETs exhibits high logic expressivity with reconfigurability at runtime to implement several key non-volatile 2-input logic functions. Moreover, the Re-FeFET circuit demonstrates remarkable compactness, with an up to 80% reduction in transistor count compared to standard CMOS design. The 2D van de Waals Re-FeFET devices therefore exhibit groundbreaking potential for both More-than-Moore and beyond-Moore future of electronics, in particular for an energy-efficient implementation of in-memory computing and machine learning hardware, due to their multifunctionality and design compactness.

**INTRODUCTION**

Modern integrated circuits face significant challenges in the quest for continuous device miniaturization, pushing the boundaries of Moore's Law.[1,2] Key concerns include device downscaling, circuit design and energy consumption. Conventional circuit design is based on elementary devices with fixed functionality such as Field Effect Transistors (FET), diodes, and memories. The prevailing technology in this domain is Complementary Metal Oxide Semiconductor (CMOS), which classically employs unipolar FETs that are fixed to either n-type or p-type operation based on their underlying fabrication process and chemical doping. However, as device downscaling progresses, this approach poses increasing complexity-related challenges.[3,4]

To overcome some of the limitations of CMOS, an elegant emerging approach leverages the concept of Reconfigurable FETs (ReFET).[5,6] Unlike conventional FETs, these devices offer a promising solution by providing the ability to adjust their functionality. A ReFET typically consists of two gate electrodes, either lateral or vertical, that independently control the energy band profile within the semiconductor channel. Specifically, the program gate (PG) determines the type of carrier, while the control gate (CG) toggles the transistor on and off. This disruptive FET concept allows for a reversible reconfiguration between n-type and p-type operating modes.[7] The inclusion of these additional functionalities in a single device enhances logic capabilities and device versatility and potentially mitigates the challenges posed by miniaturization. For instance, a logic gate based on ReFETs can dynamically adjust its configuration during operation to implement a different truth table.[8] Furthermore, this technology holds promise for next-generation hardware security applications by safeguarding against the direct reading of the layout of an integrated circuit and dramatically complicating reverse engineering through key protection strategies.[9,10] Lastly, as inherently multifunctional systems, reconfigurable electronic devices find applications in various fields such as neuromorphics[11,12], in-memory computing[13–15], energy harvesting,[16–18] and nanoelectronics in general.[19–22]

Integrating nanomaterials into the technological roadmap is a promising approach to surpass the limitations of Moore's Law.[2,5,23] Among these nanomaterials, two-dimensional (2D) materials have emerged as viable candidates due to their favorable electrical performance, scalability, and compatibility with silicon platforms.[4,24–26] The unique characteristic of 2D materials lies in their dangling bond-free interfaces, which offer unprecedented flexibility in combining different materials, thereby opening up possibilities for uncharted device concepts.[27,28] Furthermore, their ultimate atomic thickness and high sensitivity to external electric fields enable efficient engineering of the energy band profiles.[29,30] Consequently, certain 2D semiconductors, such as $WSe_2$, $MoTe_2$ or black phosphorus, can demonstrate ambipolar characteristics when used in conjunction with suitable metal contacts.[31] These combined properties position van der Waals materials as an ideal platform for reconfigurable electronics.

However, the current implementation of ReFETs heavily relies on volatile external voltages to program their operating modes. The p- and n-conduction modes are configured by connecting the program gates to fixed voltage levels.[6,9] This approach raises concerns regarding energy consumption and reliability, and limits the benefits of reconfigurability. To overcome these challenges, it is crucial to develop non-volatile ReFETs that can maintain their doping profile without the need to continuously maintain the gate voltages. One promising approach to introduce non-volatility in ReFETs is through the integration of ferroelectric materials as the gate medium. The ferroelectric gate can retain non-volatile polarization states, which are then used to electrostatically dope the adjacent semiconductor layer. This working principle is a cornerstone of the field of ferroelectronics, which encompasses various novel device concepts such as ferroelectric memories, ferroelectric tunnel junctions, photodetectors, neuromorphic circuits and ferroelectric FETs (FeFET).[32–39] In addition, novel circuit design concepts are being explored, where the internal polarization state of the FeFET serves as an input parameter for a new class of ferroelectric logic gates.[15,40–43] Recent advancements in the field have introduced ferroelectric van der Waals systems,[44–48] which include van der Waals ferroelectric heterostructures with tunable interfacial physics.[22,37,38,49–53] These discoveries have brought novel approaches into the field and opened up new possibilities. However, the exploration of reconfigurable ferroelectric electronics and their potential for Logic-in-Memory remains mostly uncharted territory.

In this work, we demonstrate reconfigurable ferroelectric devices based on van der Waals ferroelectric heterostructures. Our device architecture is based on a split-gates geometry and employs a layer of the $CuInP_2S_6$ (CIPS) van der Waals material[54,55] as the ferroelectric gate medium. This configuration enables precise, efficient, and remanent control of the energy band profile within the $WSe_2$ homojunction. Our reconfigurable Ferroelectric FET (Re-FeFET) exhibits three distinct operating electronic modes: ambipolar FeFET, unipolar n-FeFET, and unipolar p-FeFET. Furthermore, we evaluate the optoelectronic properties of the Re-FeFET. It demonstrates the capability for photodetection in both phototransistor and photodiode modes. When programmed in the photodiode mode, open-circuit voltages and short-circuit photocurrents are generated, highlighting its potential for photovoltaic applications.

Finally, we explore the versatility of these tunable electronic devices by designing reconfigurable ferroelectric logic gates, which can be programmed to implement several Boolean functions (such as NAND, NOR or XOR). We conduct a comprehensive logic gate design exploration based on structure consisting of two series-connected devices. Our results demonstrate the superiority of ferroelectric reconfigurable circuits over classical CMOS circuits. In addition, we introduce a simple model for the Re-FeFET, which facilitates further circuit design exploration and behavioral simulations, while also allowing for a deeper understanding and analysis of the device's behavior in various circuit configurations.

**RESULTS AND DISCUSSION**

**Figure 1a** depicts a schematic of the van der Waals Re-FeFET studied in this work, which is constructed from the $CIPS/hBN/WSe_2$ heterostructure. The fabrication process involves mechanically exfoliating and sequentially transferring flakes of CIPS, h-BN, and $WSe_2$ onto pre-patterned metallic back-gate electrodes. The two back-gate electrodes were patterned beforehand with a split-gate geometry and an inter-electrode spacing of 3 μm. Source and drain electrodes are then fabricated by electron-beam lithography on top of the $WSe_2$ channel. **Figure 1b** provides an optical image of a typical Re-FeFET device. To optimize the ferroelectric polarization while minimizing leakage currents, few-layered h-BN is used as a dielectric spacer. It also improves the transport characteristics of $WSe_2$ with respect to defects and trap states.[56,57] More detailed information on the fabrication process can be found in the **Supplementary Information Section 1**. In **Figure 1c**, the micro-Raman spectrum of the van der Waals heterostructure is shown, which was obtained from a typical Re-FeFET as described in the **Methods** section. The spectrum displays the first-order in-plane and out-of-plane Raman modes of the $WSe_2$ bilayer sample around 249 cm$^{-1}$ and 251 cm$^{-1}$, respectively. The recorded spectra reveal a complete suppression of degeneracy, and the two modes exhibit a frequency splitting of ~2 cm$^{-1}$. The ratio of $A_{1g}$ to $E'_{2g}$ peak intensities is about 3. The peak observed at 308 cm$^{-1}$ (labeled as $B'_{2g}$) is indicative of regions with a thickness of 2 monolayers. Based on the described Raman data, the intense $B'_{2g}$ peak confirms the bilayer behavior of the deposed WSe$_2$,[58,59] and the splitting of the two main out-of- and in-plane modes indicates a type of stacking that can be attributed to the 3R phase. Additional micro-Raman spectra, obtained from different regions of the heterostructures with varying distinct numbers of stacked van der Waals materials, are provided in **Supplementary Information Section 2**.

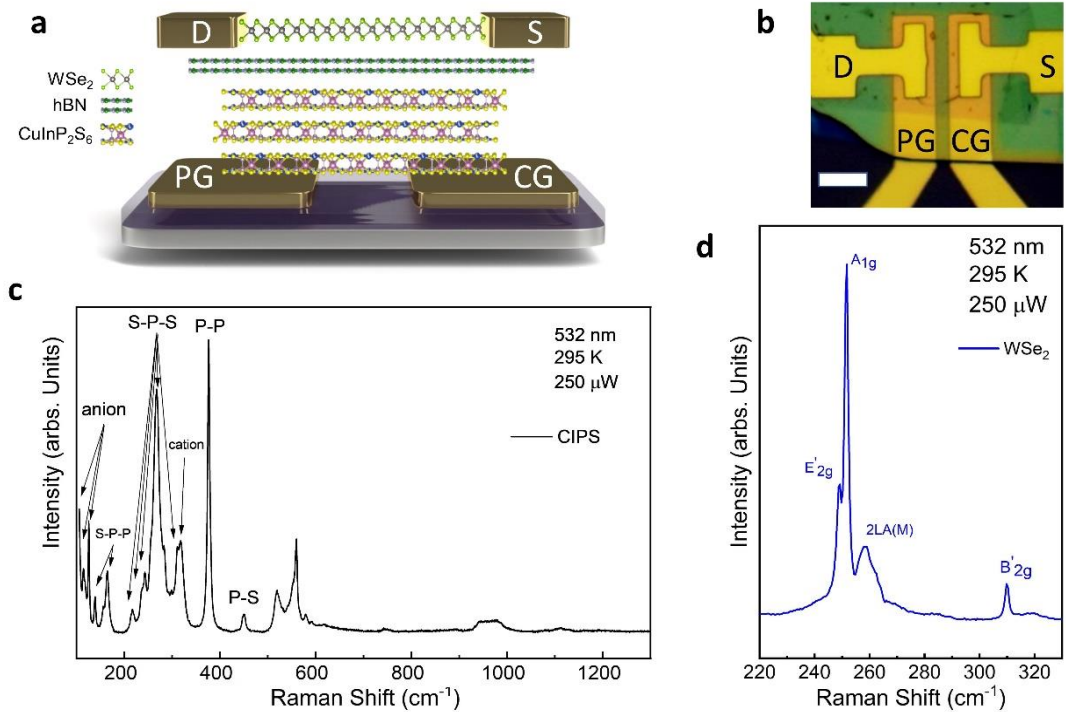

*Figure 1. Van de Waals Reconfigurable Ferroelectric Field Effect Transistor. **a**, Schematic of the Re-FeFET, with split-gate architecture combined as two independent bottom gate control terminals (Program Gate and Control Gate). The van der Waals heterostructure is composed of a WSe$_2$/h-BN/CuInP$_2$S$_6$ stack. **b**, Optical microscope image of a typical FeFET device (scale bar=10µm). **c**, Micro-Raman spectra of the CuInP$_2$S$_6$ taken from a Re-FeFET device. **d**, Micro-Raman subtracted spectra of the WSe$_2$ channel taken from the WSe$_2$/hBN/CIPS heterostructure region.*

FERROELECTRIC FIELD EFFECT TRANSISTOR WITH RECONFIGURABLE POLARITIES

We first study how the split-gate architecture makes it possible to emulate different FeFET operating behavior depending on whether the device is operated in ambipolar or unipolar mode (cf. **Figure 2a**). Transconductance measurements under different Program Gate configurations are performed while sweeping the Control Gate bias with 10 ms long voltage pulses. The **Figure 2b** shows the transconductance of the Re-FeFET operating in a symmetric configuration while biasing both gates with the same voltage (cf. "Ambipolar FeFET mode" depicted in **Figure 2a**). This operating mode is mostly equivalent to the single-gate FeFET and is well-established.[60,61] The remanent polarization state of the CIPS can be inverted by applying a gate voltage of opposite polarity, exceeding the ferroelectric coercive fields ($V_{FE}^+$ or $V_{FE}^-$ respectively for the positive or negative sides). The polarization up (P$_{up}$) or down (P$_{down}$) leads to the accumulation of negative or positive mobile charges within the WSe$_2$ channel. The Re-FeFET exhibits an ambipolar I$_{ds}$-V$_{gs}$ transfer characteristic, enabling conduction in both the electron and hole accumulated regimes. Moreover, it exhibits a well-defined hysteresis loop that can be decomposed into an anti-clockwise and a clockwise hysteresis loop respectively for the n- and p-branches. Such characteristics show that the transconductance properties of the WSe$_2$ bilayer are driven by the CIPS ferroelectric polarization switching. The analysis of the transconductance demonstrates excellent performance, with the on/off current ratio exceeding 10$^5$ (resp. 10$^4$) in the p- (resp. n-) branch, and the ferroelectric window approaching a width of 10 V. These features attest to the efficient electrostatic coupling between the CIPS and the WSe$_2$, and the good interface quality of the van der Waals stack without dominating Fermi level pinning or trap states.[62] Next, we investigate the reconfigurable properties of the Re-FeFET while operating it in asymmetric configurations (cf. the two "Unipolar FeFET modes" depicted in **Figure 2a**). Here, we firstly polarize the Program Gate either to the upward polarized state PG$^\uparrow$ (V$_{PG}$ = +12 V > $V_{FE}^+$) or to the downward polarized state PG$^\downarrow$ (V$_{PG}$ = -12 V < -$V_{FE}^-$). This enables the WSe$_2$ channel part positioned over the PG Gate to be set selectively to a predetermined doping state. Transconductance measurements are then performed while modulating only the Control Gate voltage V$_{CG}$. The Re-FeFET demonstrates now very different characteristics than those observed in the ambipolar mode: the device behaves in both asymmetric modes as unipolar-like FeFETs of opposite polarities. In the PG$^\downarrow$ state (**Figure 2b**), the Re-FeFET mimics the behavior of a p-FeFET and shows a clockwise transconductance hysteresis with an on/off current ratio of six orders of magnitude. When programmed in the PG$^\uparrow$ state, the device behaves as an n-FeFET, with a clockwise transconductance hysteresis with an on/off current ratio of four orders of magnitude (**Figure 2d**). For both unipolar modes, the high and low conductance states are obtained, respectively, when the homojunction is set to unipolar (p-p, or n-n) or ambipolar (p-n, or n-p) configurations.

To further illustrate the ambipolarity of the Re-FeFET, we perform density functional theory (DFT) calculations to find the relative alignment of the electronic energy levels of WSe$_2$ and the Ti metal contacts (see Methods for the computational details of the calculation). **Figure 2e** (left graph) depicts the WSe$_2$ and Ti slab components of the density of states (DOS) of a combined WSe$_2$-Ti system corresponding to the different states of the Re-FeFET. In **Figure 2e** (middle and right graphs), the effect of the finite surface charge of the CIPS is qualitatively illustrated by a sketch of the effective Fermi level. The WSe$_2$ and Ti slab components of the DOS are obtained by projecting the DFT orbitals onto atomic orbitals. In the absence of any surface charge in the CIPS, the Fermi level of the combined system lies in the middle of the WSe$_2$ electronic band gap (Left graph of **Figure 2e**), due to the presence of the partially filled metallic band of Ti within the gap. The choice of Ti as the contact material with its convenient level alignment to the WSe$_2$ bands thus allows for the efficient electrostatic doping of the WSe$_2$ layer with both p- and n-type carriers, leading to the achieved ambipolarity of the Re-FeFET.

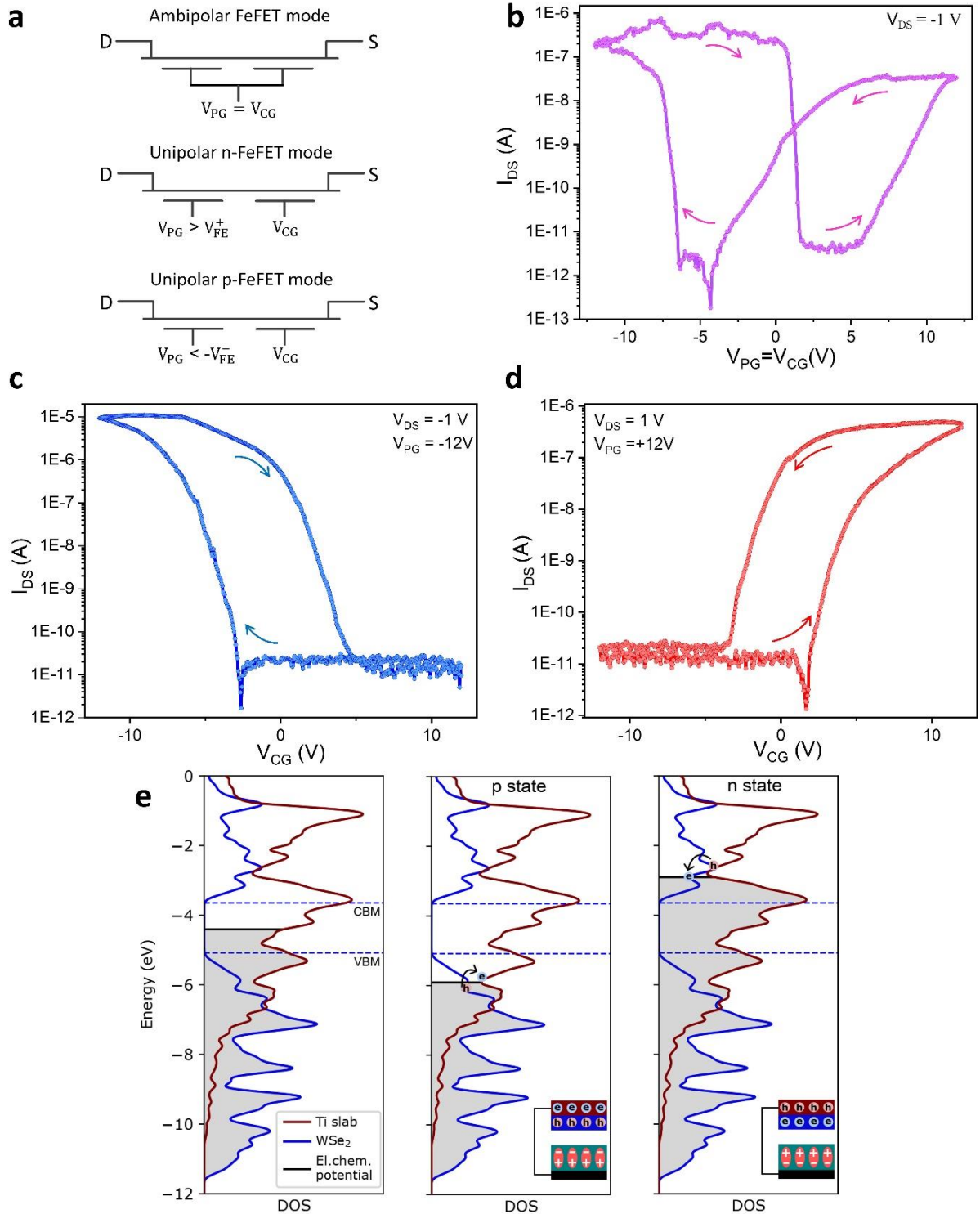

*Figure 2. Ferroelectric Field Effect Transistor with reconfigurable polarities. **a,** Schematic of the various operating modes of the Re-FeFET. In the ambipolar mode, both gates are biased at the same sweeping voltage. In the unipolar n-FeFET (resp. p-FeFET) mode, the program gate $V_{PG}$ is set above (resp. below) the positive (resp. negative) ferroelectric coercive voltage $V_{FE}$, while the control gate voltage $V_{CG}$ is swept. **b,** Transconductance loop of the Re-FeFET operating in the ambipolar mode and demonstrating ambipolar hysteresis with anticlockwise hysteresis in the n-branch and clockwise hysteresis in the p-branch. **c,** and **d,** show the transconductance loops of the Re-FeFET operating respectively in the p-FeFET and n-FeFET modes. The ferroelectric polarization switching of the CIPS is responsible for the anticlockwise (resp. clockwise) hysteresis of the n-FeFET (resp. p-FeFET) transconductance. **e,** Atom-projected density of states (DOS) of the combined system of WSe$_2$ and Ti metal slab.*

*(LEFT) The electrochemical potential lies within the Ti metallic band within the band gap of WSe$_2$. (MIDDLE) Same as (LEFT) but for the p-doped state, illustrating the electrostatic doping of the WSe$_2$ layer, with the hole charge carriers supplied by the Ti reservoir. (RIGHT) same as (MIDDLE) but for the n-doped state. Inset: Sketch of the vertical stacking of the Ti contact layer (red), the WSe$_2$ layer, and the polarized CIPS layer (teal), which is connected to the Ti through a metal contact.*

RECONFIGURABLE HOMOJUNCTION AND PHOTOVOLTAIC DIODE

We now fully explore the electrical and optoelectronic properties of the reconfigurable homojunction. We first characterize the tunability of the homojunction while investigating systematically the transport properties of the WSe$_2$ homojunction as a function of the drain-source ($V_{DS}$) and gate-source ($V_{PG}$, $V_{CG}$) bias voltages. **Figure 3a** shows the transconductance map $I_{DS}(V_{PG}, V_{CG})$ measured at fixed drain-source voltage of value 0.5V under a systematic sweep on the gate bias voltage parameters ($V_{PG}$, $V_{CG}$). Four distinct homojunction states can be identified, corresponding to the four quadrants labeled on the transconductance map: the p-p, n-n, p-n and n-p states (see corresponding energy band diagrams in **Supplementary Information Section 3**). We further characterize the rectifying capability of the homojunction with these four states and report their output characteristics in **Figure 3b**. In the unipolar states (n-n or p-p), the $I_{DS}(V_{DS})$ traces show quasi-linear behavior, while in the ambipolar states (n-p or p-n) the $I_{DS}(V_{DS})$ traces demonstrate clear non-linear rectifying characteristics. The rectification ratio (defined as the ratio of the forward current to the reverse current obtained for opposite polarities of $V_{DS}$) exceeds $10^4$ at $V_{DS}$ = 1 V in the n-p and p-n states, while it stays close to 1 in the unipolar cases (approaching respectively 1.5 and 2 for the n-n and p-p homojunctions). Such a high rectification ratio indicates the excellent quality of the p-n junction formed in WSe$_2$ and its efficient coupling with CIPS.

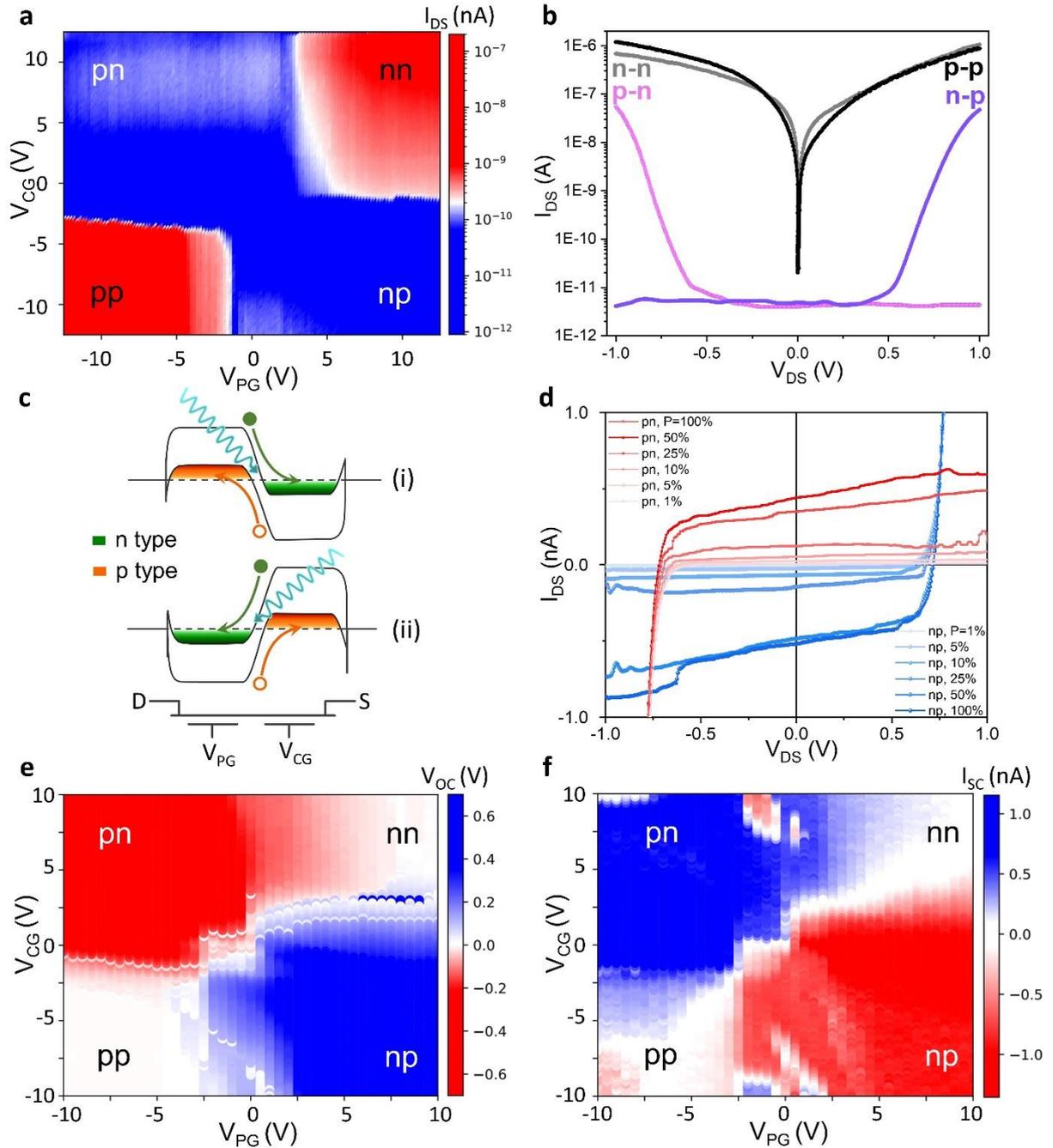

*Figure 3.* Electrical and optoelectronics tunable properties of the Re-FeFET. **a,** Contour map of the drain-source current $I_{DS}$ as a function of the voltages applied to the program gate ($V_{PG}$) and the control gate ($V_{CG}$). **b,** $I_{DS}$-$V_{DS}$ characteristics of the Re-FeFET programmed in the n-n (grey), p-p (black), p-n (pink) and n-p (purple) states. **c,** Energy band diagram illustrating the photovoltaic effect in (i) p-n and (ii) n-p homojunction configuration of the WSe$_2$ channel. **d,** Drain-source current vs voltage characteristics of the Re-FeFET programmed in p-n (red) and n-p (blue) configurations, under various light intensity "P" (P = 0.1 mW.mm$^{-2}$ at 100%, λ = 522 nm). **e,** Color map of the open circuit voltage $V_{OC}$ ($I_{DS}$ = 0 A) as the program ($V_{PG}$) and the control ($V_{CG}$) gates voltages are tuned independently. **f,** Color map of the short circuit photocurrent $I_{SC}$ ($V_{DS}$ = 0 V) as the two side-gates are swept independently.

We then investigate the potential of the Re-FeFET to convert light energy into electricity by the photovoltaic effect, taking advantage of the large built-in electric field of the p-n junction to dissociate photoexcited carriers. **Figure 3c** illustrates the separation of photogenerated electron-hole pairs due to the built-in electric field formed within the WSe$_2$ bilayer. Upon illumination, the internal built-in electric field dissociates the photogenerated carriers and gives rise to a photocurrent at zero applied voltage (short-circuit photocurrent, I$_{SC}$) and a photovoltaic voltage with zero current flow (open circuit voltage, V$_{OC}$). **Figure 3d** presents I–V curves under optical illumination ($\lambda$ = 522 nm, P = 0.001-0.1 mW.mm$^{-2}$, see **Methods** for experimental details). When the Re-FeFET is set to a diode configuration p–n (resp. n–p), the I–V characteristics are shifted upward (resp. downward) compared to the dark characteristics. This demonstrates that the Re-FeFET operates as a photodiode. **Figure 3e** and **Figure 3f** report the color maps of the I$_{SC}$ and the V$_{OC}$ respectively, as a function of the split-gate parameters (V$_{PG}$, V$_{CG}$) measured under illumination conditions ($\lambda$ = 522 nm, P = 0.1 mW.mm$^{-2}$). The color maps precisely identify the existence of short-circuit photocurrents and open-circuit voltages with the diode states previously pinpointed in dark conditions (**Figure 3a**). When the Re-FeFET is set to a unipolar state (n-n or p-p-), no I$_{SC}$ or V$_{OC}$ is detected (see **Supplementary Information Section 4**). Consistently, we observe in the p-n (n-p) configuration the apparition of a large negative (positive) V$_{OC}$ (**Figure 3d** and **Figure 3e**) and a positive (negative) I$_{SC}$ (**Figure 3d** and **Figure 3f**). The fact that I$_{SC}$ and V$_{OC}$ change sign while flipping the diode polarities is a clear indication that the photoresponse does not originate from the Schottky contacts.[17,21,63] For an optimally engineered junction profile, the maximum short-circuit photocurrent detected reaches 1.2 nA (resp. -1.3 nA) in p-n (resp. n-p) configuration, while the open circuit voltage created reaches +0.7 V (resp. -0.7 V). The photovoltaic voltages produced are among the best reported so far for WSe$_2$,[17,21,64,65] and outperform alternative van der Waals doping strategies such as superionic gating[66] and contact engineering.[67,68] This demonstrates that the Re-FeFET can also be used for energy harvesting. Moreover, the non-volatile ferroelectric control of the photodiode properties opens new prospects for in-memory sensing and computing.[69]

POLYMORPHIC FERROELECTRIC AND ELECTRONIC LOGIC GATES

As mentioned before, the Re-FeFET architecture enables the programming of distinct homojunction states with different conduction levels. This naturally hints at reconfigurable Boolean logic operations. The schematic of **Figure 4a** shows a simple logic circuit built from a Re-FeFET in series with a passive load. We now investigate the reconfigurability of this logic circuit while operating it either as an electronic or a ferroelectric logic gate. In the *electronic mode*, the parameters (V$_{CG}$, V$_{PG}$) define the two input variables (schematic of **Figure 4a**). Here, V$_{CG}$ and V$_{PG}$ are set independently to the "1" state (resp. the "0" state) while polarizing the CIPS to the upward (resp. downward) polarization with V$_{CG}$ or V$_{PG}$ > $V_{FE}^{+}$ (resp. V$_{CG}$ or V$_{PG}$ < -$V_{FE}^{-}$). As shown in the measured output voltage V$_{out}$ in **Figure 4a**, the Re-FeFET circuit operated in the electronic logic mode allows emulation of the XNOR logic function. In the case of the *ferroelectric mode*, the logic state is now set by the input pair (FE$_{Branch}$, V$_{CG}$). The program gate (V$_{PG}$) is pre-set to a fixed polarized state (either in the PG$^{\uparrow}$ state or the PG$^{\downarrow}$ state) as detailed in the schematics of **Figure 4b** and **Figure 4c**. The new variable FE$_{Branch}$ is specific to ferroelectric logic.[15,43] It is set to the "0" state (resp. the "1" state) when the FeFET is operating in the low (resp. high) threshold voltage branch of the transconductance hysteresis that transits at the negative (resp. positive) coercive field $V_{FE}^{-}$ (resp. $V_{FE}^{+}$). As shown in **Figure 4b** and **Figure 4c** (see also **Supplementary Information Section 5**), the Re-FeFET logic can switch between a ferroelectric NAND (FE-NAND) and a ferroelectric AND (FE-AND) function respectively, depending on the pre-programmed FeFET state (either p- or n- type, respectively). Hence, various ferroelectric logic functions can be obtained, providing polymorphic logic functionality, which is of interest for hardware security, custom operations, and circuit design. It is important to underline that the logic gate inputs can be provided by voltage pulses. In this case, the $I_{DS}$ current is measured after applying the various gate pulses, in open-circuit gate conditions, and the ferroelectric logic states correspond to remanent states. The retention properties of the four homojunction states are reported in **Figure 4d** and **Figure 4e**. The Program and Control Gates have been programmed independently either in the up or down state, and subsequently grounded. The output voltage is presented as a function of time, and demonstrates successful retention after more than 10^4 seconds for each state, comparable to the best values reported so far on van der Waals ferroelectric heterostructures.[70] Such remanence is a major added value when compared with traditional CMOS-based circuits that rely on a constant supply of external voltages to maintain the desired doping levels. This represents a paradigm shift with respect to conventional von Neumann architectures, as the information processing and storing are combined in a single device. This demonstrates the promise of van der Waals ferroelectric circuits for Logic-In-Memory hardware.[15]

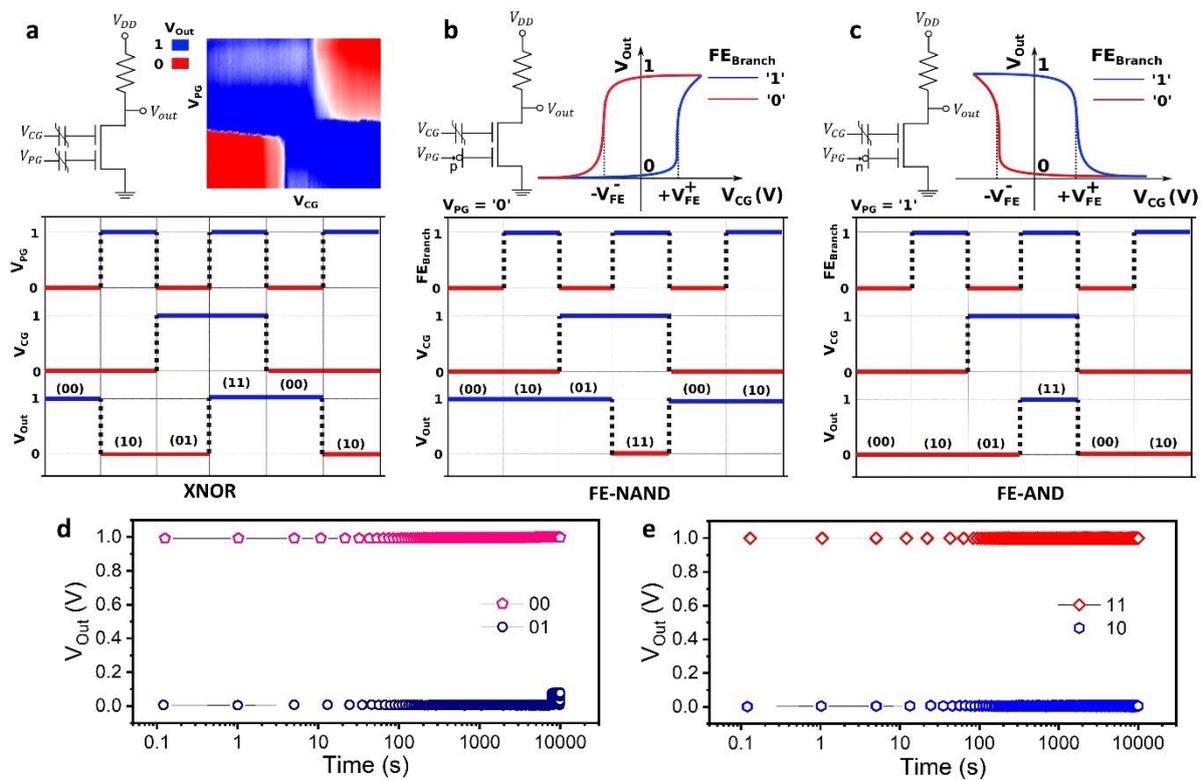

*Figure 4.* Polymorphic ferroelectric logic gate. **a,** Ferroelectric circuit (cf. circuit schematic) based on Re-FeFET operating in ambipolar mode, and demonstrating XNOR logic. **b,** Same circuit with the Re-FeFET programmed to the n-FeFET mode, demonstrating NAND logic. **c,** Same circuit with the Re-FeFET operating in the p-FeFET mode and showing AND logic. When used as a polymorphic ferroelectric logic gate (**a,b**), both inputs represent the in-memory state corresponding to the FE$_{Branch}$ (cf. top schemes V$_{out}$(V$_{CG}$) hysteresis traces in **a,** and **b,** and the V$_{CG}$ voltage (state '1' for $V_{FE}^+ > V_{CG} > 0$ V, state '0' for $V_{CG} > V_{FE}^+$). When operating as an electric logic gate, both inputs are provided by V$_{CG}$ and V$_{PG}$ terminals (cf. color plot scheme in **c,**). Waveforms V$_{out}$ vs. logic state are traced based on experimental data, with V$_{DD}$ = 1 V. **d, and e,** show the retention properties of the four ferroelectric logic states of the Fe-XNOR gate. The output voltage (normalized respect to V$_{DD}$ set at 0.5 Volt) was measured after the Program and Control Gates have been programmed independently either in the up or down state, and subsequently grounded.

## CIRCUIT DESIGN EXPLORATION WITH RE-FEFETS

In this section, we assess the potential of ferroelectric reconfigurable logic circuits based on Re-FeFETs through a circuit design exploration and modeling study. In order to set a framework and terminology for the analysis, we abstract the single device to a set of parameters $\{\{t, PG\}, \{X, CG\}\}$ where:

1. $t$ represents the non-volatile configured type of the transistor, where $t = 0$ indicates p-type operation and $t = 1$ indicates n-type operation.
2. $PG$ represents the volatile voltage applied to the program gate (also referred as polarity gate) after type configuration, causing a shift of the threshold voltage.
3. $X$ represents the non-volatile polarization of the second part of the ferroelectric layer within the Re-FeFET. Here, $X = 0$ (resp. 1) corresponds to a negative (resp. positive) polarization.
4. $CG$ represents the volatile voltage applied to the control gate after polarization configuration.

This parameter set for the Re-FeFET allows us to establish conditions that determine when the device is in the $On$ state or the $Off$ state. These conditions are summarized in the conduction table, as presented in **Figure 5a**. For more detailed explanations regarding the axioms and underlying principles used to construct this conduction table, the reader is referred to the **Supplementary Information Section 6**.

With this formalization of the Re-FeFET functionality, it becomes possible to explore the diverse logical functionalities that can be achieved with simple circuit structures. Let us consider a specific case where the circuit consists of two series-connected Re-FeFETs (shown in **Figure 5c**). By leveraging the conduction table, non-volatile logic operations can be implemented using this circuit. The process involves searching for solutions within the table that satisfy the constraints imposed by the truth table of the desired logical operation. Each solution corresponds to a unique parameter set for each Re-FeFET in the circuit ($Re\text{-}FeFET_1 = \{\{t_1, PG_1\}\{X_1, CG_1\}\}$ and $Re\text{-}FeFET_2 = \{\{t_2, PG_2\}\{X_2, CG_2\}\}$).

To create any logical or computing function, six elementary logic gates are needed: $NAND, NOR, AND, OR, NOT$ and $XOR$ The conduction table provides direct implementations for $NAND$ and $NOR$ functions. However, the $XOR$ function requires a more complex implementation. Detailed information on the implementation of $NAND, NOR$, and $XOR$ functions are given in the **Supplementary Information Section 6**. From the $XOR$ truth table, it is possible to find solutions within the conduction table by setting $n_1 = V_{dd}$ and $n_2 = gnd$, given by:

- $Re\text{-}FEFET_1 \in \{\{\{\bar{B}, 0\}, \{1, A\}\}, \{\{\bar{B}, A\}, \{0, 1\}\}, \{\{\bar{B}, A\}, \{1, 0\}\}\}$,
- $Re\text{-}FEFET_2 \in \{\{\{B, 0\}, \{1, A\}\}, \{\{B, A\}, \{0, 1\}\}, \{\{B, A\}, \{1, 0\}\}\}$.

It is worth noting that these solutions only utilize $A$ as the volatile operand. This implies that there is no need for an additional inverter to physically implement this circuit, unlike conventional implementations of 2-input XOR gates. Additionally, the non-volatile operand $B$ is linked to the *type* of the Re-FeFET. Consequently, $B$ and $\bar{B}$ are set within each Re-FeFET using an external shared programming circuit and it is not necessary to generate $\bar{B}$ with a local inverter. In some cases, when an n-type device is used to pass $V_{dd}$ to the output, there can be a logic '1' degradation. To mitigate logic degradation, it is common practice to regenerate the output voltage level (usually via an inverter and/or a level restorer), either for each logic gate or for a set of cascaded logic gates.

Finally, it is demonstrated that all six elementary logic gates can be accessed using the single series-connected Re-FeFET (**Figure 5c**) circuit configuration, while keeping $n_1 = V_{dd}$ and $n_2 = gnd$. An example configuration for each logic gate implementation is depicted in **Figure 5b**. As shown in **Figure 5**, Re-FeFETs offer the advantage of very compact logic gates compared to CMOS circuits. In the case of the inverter, the same transistor-count is required for both Re-FeFET and CMOS implementations. However, **Figure 5d** and **Figure 5e** show that for $NAND$ and $NOR$ gates, the Re-FeFET based circuit requires only two devices, while the equivalent CMOS circuits necessitate four transistors. In the case of the $XOR$ operation presented in **Figure 5f**, the CMOS implementation typically requires 12 transistors, considering the two additional inverters required to convert $A$ and $B$ to their complemented forms ($\bar{A}$ and $\bar{B}$). On the other hand, the Re-FeFET based $XOR$ can be implemented with just two devices. This results in a significant reduction in transistor count by a factor of 6. Consequently, Re-FeFET devices offer the potential for compact circuits, which can be beneficial in a wide range of applications such as reconfigurable circuits with systematic and regular structures (such as FPGAs), or obfuscated circuits aimed at preventing reverse engineering, among others. The use of an identical basic circuit element for all logic gates provides additional advantages. Fabrication constraints are alleviated, leading to more compact and reliable chip designs. Furthermore, hardware security can be enhanced as the time-dependent power consumption of each logic gate remains identical, making it more challenging to derive information from power measurements. This inherent resilience to side channel attacks adds to the appeal of Re-FeFET-based circuits in terms of hardware security.

To facilitate further design considerations and enable device simulation, an equivalent circuit is created in the *Cadence* software. The details of this circuit can be found in the **Supplementary Information Figure S5**. Using this equivalent circuit, behavioral simulations of the three main logical operations ($NAND$, $NOR$, and $XOR$) are conducted. The simulation results are presented in **Figure 5g**.

**A** Re-FeFET conduction table

|   |    | $t=0$ (p-type) |        | $t=1$ (n-type) |        |
|---|----|----------------|--------|----------------|--------|
| X | CG | $PG=0$         | $PG=1$ | $PG=0$         | $PG=1$ |
| 0 | 0  | On             | On     | Off            | Off    |
| 0 | 1  | On             | Off    | Off            | On     |
| 1 | 0  | On             | Off    | Off            | On     |
| 1 | 1  | Off            | Off    | On             | On     |

**B** Re-FeFET serial circuit configuration for elementary logic gates implementation

|           | Re-FeFET$_1$ |        |             |             | Re-FeFET$_2$ |        |             |             |
|-----------|--------------|--------|-------------|-------------|--------------|--------|-------------|-------------|
| Operation | $t_1$        | $PG_1$ | $X_1$       | $CG_1$      | $t_2$        | $PG_2$ | $X_2$       | $CG_2$      |
| NOT       | 0            | 1      | 0           | A           | 1            | 1      | 0           | A           |
| NAND      | 0            | 0      | B           | A           | 1            | 0      | B           | A           |
| OR        | 0            | 0      | $\bar{B}$   | $\bar{A}$   | 1            | 0      | $\bar{B}$   | $\bar{A}$   |
| NOR       | 0            | 1      | B           | A           | 1            | 1      | B           | A           |
| AND       | 0            | 1      | $\bar{B}$   | $\bar{A}$   | 1            | 1      | $\bar{B}$   | $\bar{A}$   |
| XOR       | $\bar{B}$    | A      | 1           | 0           | B            | A      | 1           | 0           |

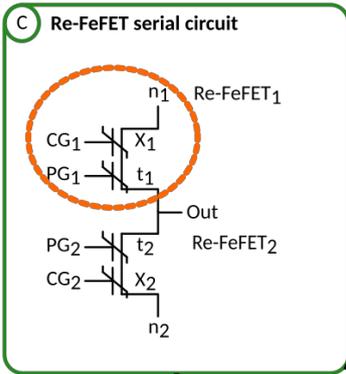

**C** Re-FeFET serial circuit

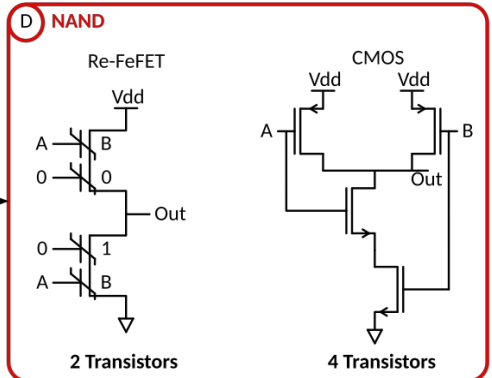

**D** NAND — Re-FeFET: 2 Transistors — CMOS: 4 Transistors

Factor 2X

Factor 6X

Factor 2X

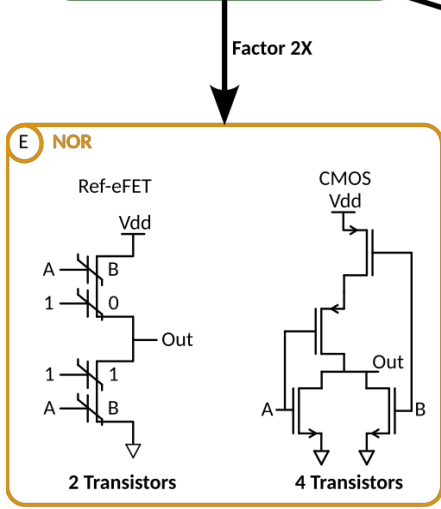

**E** NOR — Ref-eFET: 2 Transistors — CMOS: 4 Transistors

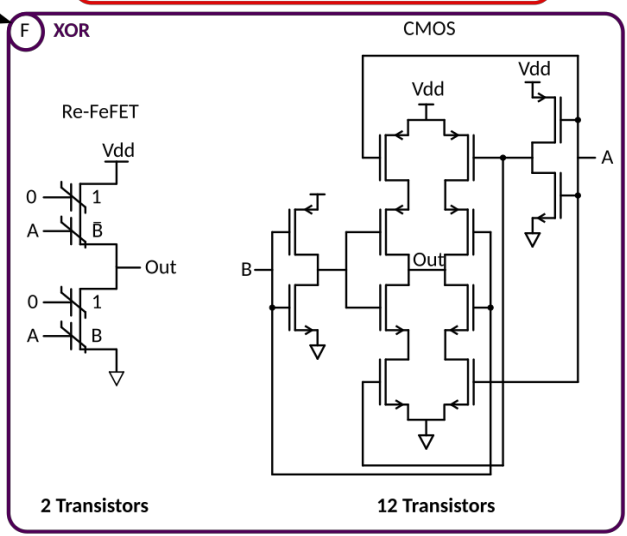

**F** XOR — Re-FeFET: 2 Transistors — CMOS: 12 Transistors

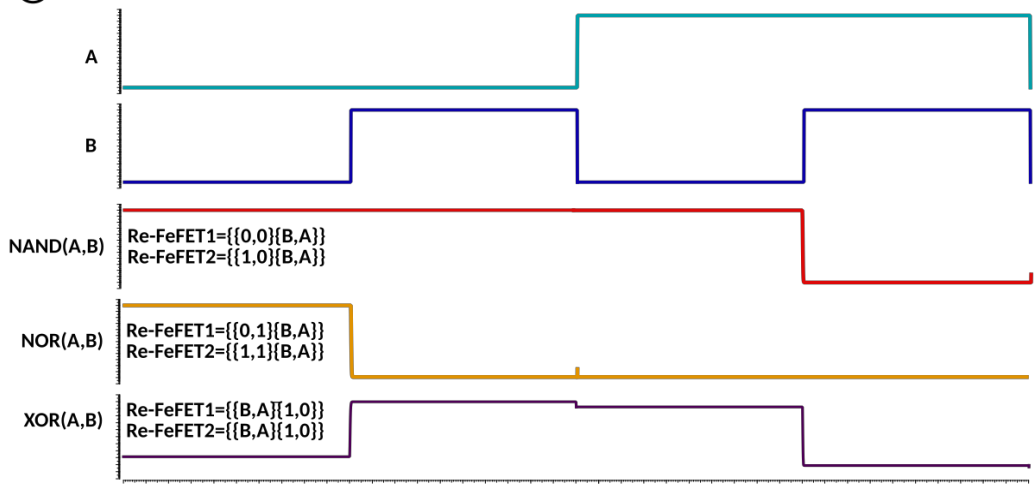

**G** Cadence simulation of the equivalent circuit for the 2-input NAND, NOR, and XOR

NAND(A,B): Re-FeFET1={{0,0}{B,A}}, Re-FeFET2={{1,0}{B,A}}
NOR(A,B): Re-FeFET1={{0,1}{B,A}}, Re-FeFET2={{1,1}{B,A}}
XOR(A,B): Re-FeFET1={{B,A}{1,0}}, Re-FeFET2={{B,A}{1,0}}

*Figure 5. Implementation of NAND, NOR and XOR functions using Re-FeFETs as compared to conventional CMOS circuits. **a,** Re-FeFET conduction table specifying when the Re-FeFET is conducting ("On") or not ("Off") depending on the set of parameters. **b,** Table of series-connected Re-FeFET circuit (**c,**) configurations for six different logic functions (NOT, NAND, OR, NOR, AND and XOR). **c,** Ferroelectric circuit consisting of two series-connected Re-FeFETs and detailed configurations to the logic functions NAND (**d,**), NOR (**e,**), and XOR (**f,**). For each logic function, the ferroelectric logic demonstrate superiority in terms of circuit compactness with a factor between 2 and 6. **f,** Behavioral simulations (created with the Cadence software) of the three main logic operations (NAND, NOR, XOR) performed by the in-series Re-FeFET circuit.*

## CONCLUSION

In conclusion, our exploration of reconfigurable ferroelectric devices based on a $WSe_2$ semiconducting channel controlled by split-gates and the $CuInP_2S_6$ van der Waals material has revealed significant properties and functionalities. We demonstrate full electrostatic control of the device by means of the split-gate electrodes. This architecture enables the precise tuning of the energy band profile along the $WSe_2$ channel from unipolar (n-n or p-p) to ambipolar (n-p or p-n) homojunction configurations. The Re-FeFET can be switched between p-FeFET, n-FeFET, and ambipolar-FeFET modes. Using DFT-level electronic structure calculations, we showed that the favorable alignment of the Fermi level of the Ti contacts within the band gap of the $WSe_2$ layer allows its electrostatic doping with both p-type and n-type carriers. With careful device engineering, the Re-FeFET demonstrates an on-off ratio of $10^6$ ($10^4$) for the p-branch (n-branch), clear carrier-polarity clockwise/anticlockwise reversal of ferroelectric hysteresis and a wide memory window of up to 10 V. Additionally, the Re-FeFET can be configured as a diode (pn or np) to provide photovoltaic functionalities. The diode homojunction configurations demonstrate an excellent rectification ratio exceeding 4 order of magnitude. The observed photocurrent and large open-circuit photovoltage are attributed to the efficient electron-hole separation facilitated by the built-in electric field at the PN junction. Furthermore, we have demonstrated a proof-of-concept reconfigurable logic unit that can function either as a polymorphic ferroelectric logic or as an electronic logic unit. In the ferroelectric mode, the reconfigurable logic unit can be switched from a Fe-NAND gate to a Fe-AND gate based on the state stored in the Program Gate, and demonstrate long retention capability exceeding 10^4 s. Such promising characteristics call for future works to explore the endurance of Re-FeFET for data-intensive applications and the effect of device downsizing. When operated in the electronic mode, the independent control of the Program and Control Gates allows for encoding the XNOR logic. We have also explored the further potential of the Re-FeFET for reconfigurable logic circuit design purposes by first formalizing the device as a set of parameters, then proposing a structure composed of two series-connected Re-FeFETs. By applying the formalized parameter set, we have analyzed various configurations of the structure to implement 2-input NOT, NAND, OR, NOR, AND, and XOR logic functions.

Our findings demonstrate that the Re-FeFET based structure enables significantly more compact circuit implementations compared to conventional CMOS circuit implementations, resulting in up to 80% reduction in transistor count. However, it will be necessary to reduce the voltages used to program and operate the Re-FeFET in future work in order to compare fairly with conventional CMOS platforms. Identified pathways include reducing the thickness of the ferroelectric layer[46,71] and contact engineering. [67,68] Nevertheless, the high logic expressivity enabled by the simple Re-FeFET-based circuit under study will help to reduce complexity and increase logic circuit compactness through reconfigurability. We have also validated the behavior of our logic gate using equivalent circuit-based simulations, further confirming the feasibility and performance of the Re-FeFET in logic circuit applications. This work thus establishes the foundations for future research on reconfigurable ferroelectric devices and circuits, showcasing their potential in logic functions, reconfigurable circuits, in-memory computing, and photodetection. Our proof-of-concept device is thus a step toward the fabrication of non-volatile reconfigurable electronic circuits. We envision that ferroelectric circuits represent a large playground for heterostructure engineering with 2D materials and for nanoelectronics systems design. Hence, while the gate overlap scheme was engineered in our device, the Re-FeFET operating mechanism is also fully compatible with the gate underlap geometry. This comes with anticipated added benefices in terms of device performance and upscaling.[72] Also, alternative ambipolar van der Waals semiconductors with smaller bandgap energy, such as black phosphorus or molybdenum ditelluride, should be investigated with expected smaller ferroelectric operating window.[31] Furthermore, the fascinating van der Waals sliding ferroelectrics are promising candidates to enable the ferroelectric gate control down to thin layer of just few atoms.[73,74] Finally, our results highlights the promising nature of van-der-Waals materials as a platform for engineering high-performance beyond-Moore devices. The ability of the Re-FeFET to offer multiple functionalities in a single unit (electronic, memory, optoelectronic, and photovoltaic), with reconfigurability at runtime, holds promise for both More-than-Moore and beyond-Moore strategies in the field of advanced electronics.

**METHODS/EXPERIMENTAL**

**Optical spectroscopy measurements**

Micro-Raman spectroscopy measurements were conducted at room temperature using a commercial confocal Horiba Labram Raman spectrometer operating with a laser of wavelength λ = 532 nm. The laser beam was focused onto a small spot with diameter of about 1 μm on the sample and incident power of about 250 μW. All measurements are acquired at room temperature.

**Opto-electrical measurements**

The high precision source meters K2634B were used for electrical characterization and gate pulse control. For the ferroelectric logic circuit, the K2182A nanovoltmeter was used to extract the voltage output parameter, while measuring the voltage drop across a resistance connected in series with the device. Voltage pulses, or combination of voltage pulses and voltage sweeps, were applied to the PG (Program Gate) and CG (Control Gate) inputs to produce different states to characterize the different gate logics (XNOR, NAND, AND). In-situ optical fiber was used for photoexcitation together with COHERENT OBIS LS/LX 522 nm source in CW-power mode.

**Circuit modeling and simulation**

Circuit modeling and simulation are performed using an equivalent circuit implemented in Cadence software for behavioral simulations.

**DFT modeling**

The DFT calculations for the level alignment of WSe$_2$ and the Ti contact are performed with the *PWscf* program of the *QuantumESPRESSO* suite[75,76]. A plane-wave basis for the single-electron wave functions is used, truncated at a kinetic energy cutoff of 60 Ry. The interaction of the ionic core with the valence electrons is modeled using ultrasoft pseudopotentials from the PS library[77]. The plane-wave basis for the electronic density is truncated at 540 Ry. The exchange and correlation interactions are described using the Perdew-Burke-Ernzerhof parametrization of the generalized gradient approximation.[78] The spin-orbit interaction is neglected in the interest of computational efficiency, as the density of states of WSe$_2$ is only marginally effected. The structures of the individual materials were fully optimized, yielding in-plane lattice constants of 3.32 Å and 2.93 Å for WSe$_2$ and Ti, respectively.

To get the absolute level alignment with respect to a common vacuum level, a commensurate hexagonal supercell of a monolayer WSe$_2$ ($\sqrt{7} \times \sqrt{7}$), rotated by 18°, and a 10-atom thick Ti (0001) slab (3 × 3) is constructed using the CellMatch package[79] to achieve a minimal strain of 0.01%. As we are interested in finding the relative electronic level alignment of the metal and freestanding 2D semiconductor, rather than the complex potential landscape at their interface,[80,81] the WSe$_2$ layer and Ti slab are separated in our simulation by a vacuum distance of 12.5 Å. For integrations over the first Brillouin zone, we use a non-shifted Monkhorst-Pack grid of 6×6×1 points. We use a Gaussian smearing for the metallic Ti states with a smearing energy of 50 meV. The Ti slab is structurally relaxed to obtain the ground state geometry of the surface. The band energies obtained using DFT are aligned by setting the electrostatic potential (ionic+Hartree) to zero at a point in the vacuum region of the supercell, equidistant from the WSe$_2$ layer and the Ti slab.

**SUPPORTING INFORMATION**

Device Fabrication and AFM characterization; Raman characterization; Reconfigurable Energy band profiles; Reconfigurable FeFET operating in Phototransistor mode; Ferroelectric logic gate measurements; Circuit design exploration with Re-FeFETs.

**AUTHOR CONTRIBUTIONS**

J.F.D. planned the experiments and supervised the project. A.R. fabricated the samples, lead and performed electrical measurements under the supervision of J.F.D. C.M. and I.O.C developed the ferroelectric circuit design and simulations. K.M set-up the electrical measurement bench under supervision of J.F.D and B.D. A.M. and A.O. did Raman analysis. A.R.K. and S.R. performed the first-principles calculations. T.T. and K.W. provided *h*-BN crystals. M.S. contributed to sample fabrication and electrical measurements. A.R., K.M., C.M., I.O.C. and J.F.D. analyzed the experimental and circuit design data. All authors discussed the results and their implications and commented on the manuscript. J.F.D., I.O.C and C.M. supervised the manuscript writing.

**ACKNOWLEDGMENTS**

We also acknowledge the Agence Nationale de la Recherche for financial support through the grant MixDFerro (ANR-21-CE09-0029) and MIXES (ANR-19-CE09-028), the Region Grand-Est (Project Phoenix), IdEx Unistra (ANR 10 IDEX 0002), SFRI STRAT'US project (ANR 20 SFRI 0012) and EUR (QMat-ANR-18-EUR-0016) under the framework of the French Investments for the Future Program. We acknowledge support from Eucor - The European Campus for financial support through the QUSTEC funding from the European Union's Horizon 2020 research and innovation program under the Marie Sklodowska-Curie Grant Agreement No. 847471, and support of the Institut Universitaire de France (IUF). A.R.K. and S.R. acknowledge funding from the National Research Fund (FNR) Luxembourg, project « RESRAMAN » (grant no. C20/MS/14802965). We thank Bohdan Kundys for providing diodes and his techinal advise on optical measurements, Fabien Chevrier for cryogenic technical assistance, and the STnano nanofabrication platform.

**Supporting Information**

# Reconfigurable Multifunctional van der Waals Ferroelectric Devices and Logic Circuits

Ankita Ram[1], Krishna Maity[1], Cédric Marchand[2], Aymen Mahmoudi[3], Aseem Rajan Kshirsagar[4], Mohamed Soliman[1], Takashi Taniguchi[5], Kenji Watanabe[5], Bernard Doudin[1,6], Abdelkarim Ouerghi[3], Sven Reichardt[4], Ian O'Connor[2]* and Jean-Francois Dayen[1,6]*.

1. Université de Strasbourg, IPCMS-CNRS UMR 7504, 23 Rue du Loess, 67034 Strasbourg, France.
2. École Centrale de Lyon, 36 Avenue Guy de Collongue, Ecully 69134, France.
3. Université Paris-Saclay, CNRS, Centre de Nanosciences et de Nanotechnologies, 91120, Palaiseau, France.
4. Department of Physics and Materials Science, University of Luxembourg, Luxembourg 1511, Luxembourg.
5. National Institute for Materials Science, Tsukuba, Ibaraki 305-0044, Japan.
6. Institut Universitaire de France, 1 rue Descartes, 75231 Paris cedex 05, France

# Contents

# Section 1. Device Fabrication

The fabrication process of the device started with the patterning of the bottom split-gate electrode on a silicon/silicon dioxide (Si/SiO$_2$) substrate using electron beam lithography (EBL). A layer of poly methyl methacrylate (PMMA) resist was spin-coated onto the substrate, followed by EBL exposure using a Zeiss Supra 40 system and Raith Elphy Plus software. The resist was subsequently developed in a 1:3 mixture of methyl isobutyl ketone and isopropyl alcohol. The patterned substrate was then subjected to e-gun evaporation, depositing a 5 nm titanium layer followed by a 45 nm gold layer with controlled thickness and uniformity. Lift-off was performed to remove the excess metal, yielding a precisely defined bottom gate electrode, see **Figure S1.1a(2)** for schematic and **Figure S1.1b(2.i)** for optical image.

Next, using the scotch tape exfoliation technique, van der Waals materials have been exfoliated from their bulk 3D crystals and transferred onto a PDMS polymer fixed on a glass slide. After exfoliation, a careful optical microscope inspection facilitated the identification and isolation of flakes with desired qualities and suitable thickness (**Figure S1.1.b(1)**). We have done 3 different Re-FeFET samples, with typical materials thicknesses (see **Figure S.1.2.a-d**): CIPS(90 nm, 200 nm, 380nm), hBN (8 nm, 9 nm, 10 nm) and WSe$_2$ (2 nm, 3.5 nm, 4 nm). We estimate the height error, considering surface roughness and possible interlayer water, in the range of ±1 Layer for WSe$_2$.[1,2]

To achieve the desired stack, a lab-built micro positioner transfer station was used for precise positioning and layer-by-layer transfer of the flakes onto the previously patterned bottom gate electrode. This process involved heating the substrate to 45°C, allowing for controlled expansion of the PDMS stamp upon contact. The stamp was gradually brought into contact with the substrate, with the temperature incrementally raised by 10 degrees to ensure full contact. The PDMS retraction was carefully executed using mechanical and temperature control to avoid any damage or misalignment. This layer stacking process was repeated for each individual layer, resulting in the formation of the desired stack of layered materials, see **Figure S1.1.a(3-5)** for schematic and **Figure S1.1.b(2.ii-iv)** for optical images.

Subsequently, the top electrodes patterning was accomplished through EBL, following the same lithography procedure as employed for the bottom gate electrode but with enhanced alignment techniques. The top surface of the layered stack was patterned using EBL exposure, and a 10 nm titanium layer along with a 150 nm gold layer was deposited through e-gun evaporation. The deposition of a thicker Ti/Au top contact aimed to prevent any cracks that may occur due to increased thickness of the entire 2D stack. Finally, lift-off was performed to remove the excessive metal, defining the well-aligned top electrodes, see **Figure S1.1.a(6)** for schematic and **Figure S1.1.b(2.v)** for optical image. Channel length is set to 6 µm, PG to CG gap distance is 3 µm, while the channel width is 15 µm large.

The device fabrication was completed by performing manual bonding using silver paste and epoxy to establish electrical connections with gold wires. This ensured reliable electrical connections between the device and external measurement equipment. Subsequently, the bonded device was mounted in a cryostat, providing a controlled environment for electrical measurements.

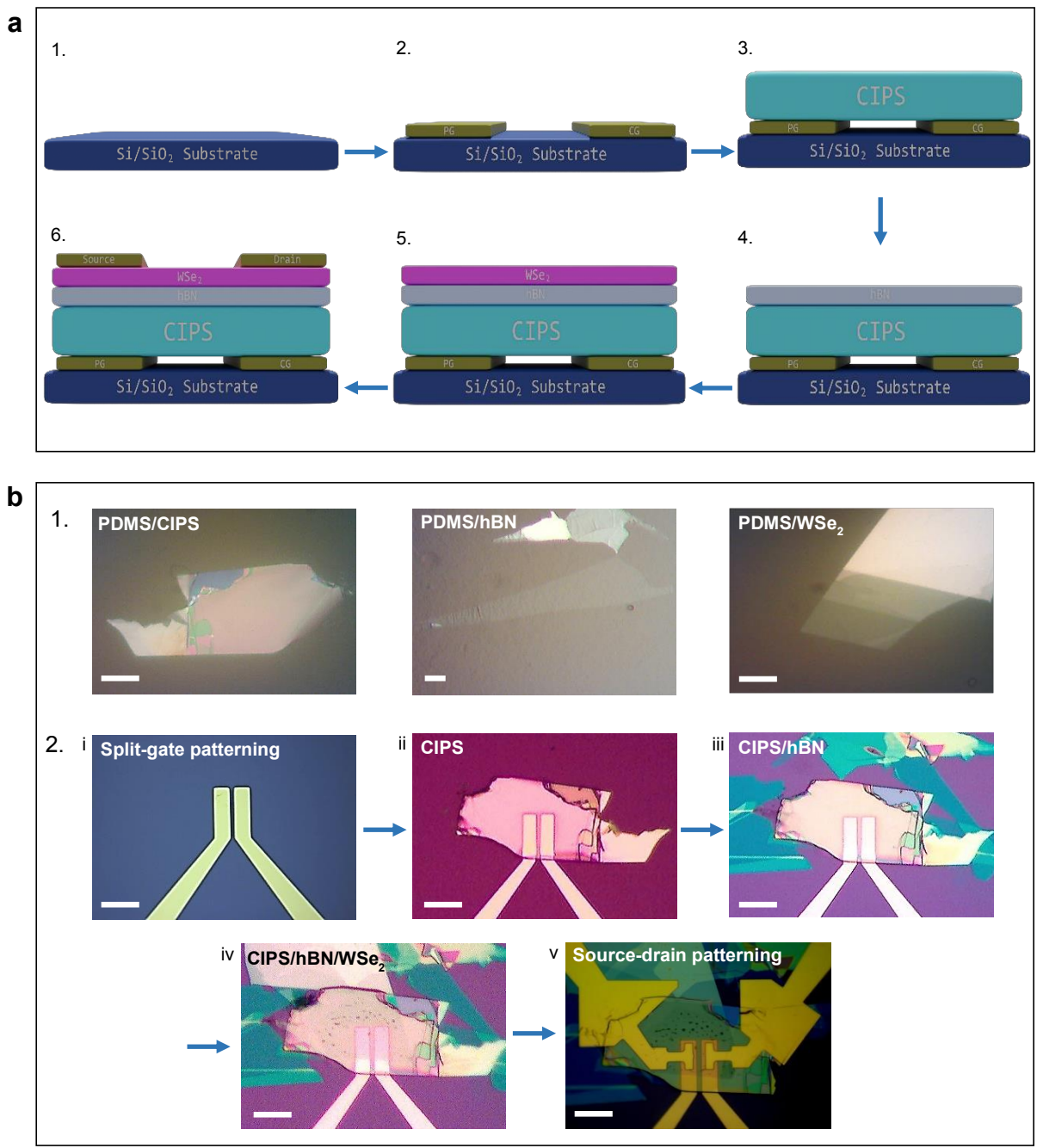

*Figure S1.1 Device Fabrication process main steps. **a**, Schematic diagram of process flow and layer-by-layer assembling. **b**, Optical microscope images (scale bar = 20 μm) of 1. exfoliated CIPS, hBN and WSe₂ flakes on independent PDMS substrates, 2. different fabrication steps of a typical Re-FeFET (from bottom-gate patterning, layer-by-layer stacking, to top electrodes pattering).*

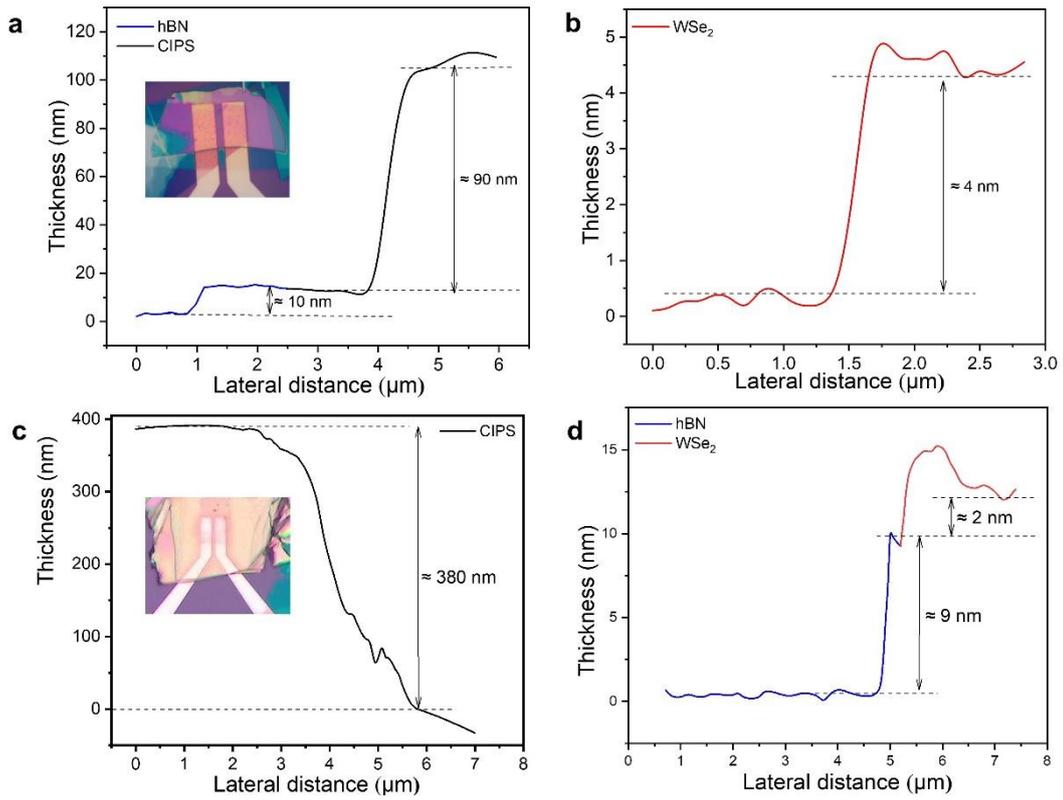

*Figure S1.2 **a-d,** Height profiles of hBN, CIPS and WSe$_2$ measured on Re-FeFET devices by Atomic Force Microscopy. Inset of **a** (resp. **c**) shows optical microscope images of sample characterized in **a** and **b** (resp. **c** and **d**)*

## Section 2. Raman characterization

**Figure S2.a** shows micro-Raman spectra taken from different regions of the heterostructures, with distinct numbers of van der Waals materials stacked : WSe$_2$/hBN/CIPS, hBN/CIPS, and CIPS regions. The measured Raman response from the naked CIPS is analogous to the previous measurement taken from a similar device in our later work.[3] For the spectrum from the hBN region, the hallmark peak of hBN is appearing around 1365 cm$^{-1}$.[4,5] **Figure S2.b** shows additional spectra of the WSe$_2$/hBN/CIPS region for the sake of statistical analysis.

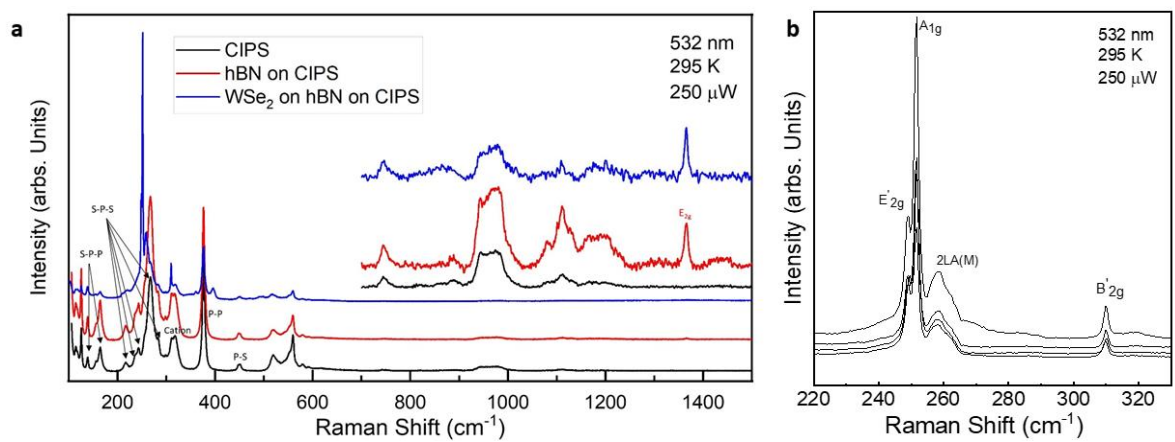

*Figure S2. **a**, Micro-Raman spectra taken from WSe$_2$/hBN/CIPS, hBN/CIPS, and CIPS regions. **b**, More recorded spectra on the WSe$_2$/hBN/CIPS region providing a statistical analysis to confirm the response behavior.*

## Section 3. Reconfigurable Energy band profiles

The energy band profile of the WSe$_2$ semiconductor channel can be reconfigured thanks to the two independent side gates. **Figure S3** shows the energy band diagrams of the Re-FeFET operated under various ($V_{PG}$, $V_{CG}$) settings, enabling to encode unipolar or ambipolar homojunction profiles.

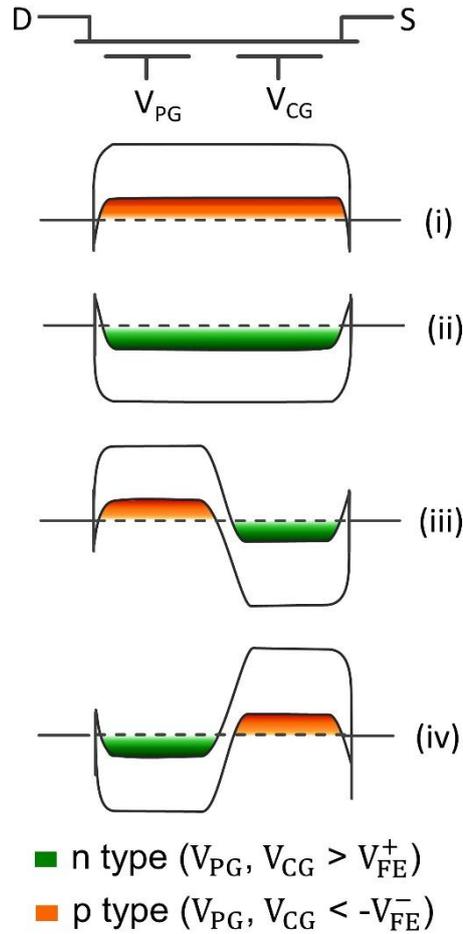

*Figure S3. Energy band diagrams of the Re-FeFET operated under various ($V_{PG}$, $V_{CG}$) settings, respectively in the (i) p-p, (ii) n-n, (iii) p-n and (iv) n-p modes.*

# Section 4. Reconfigurable FeFET in Phototransistor mode

**Figure S4** presents the output characteristics of the drain-source current ($I_{DS}$) plotted against the voltage ($V_{DS}$) for the unipolar states (n-n and p-p), in the dark and under illuminated conditions (P = 0.1 mW.mm$^{-2}$). Notably, the $I_{DS}$-$V_{DS}$ curves intersect at the origin, indicating the absence of both short-circuit current ($I_{SC}$) and open-circuit voltage ($V_{OC}$). This demonstrates that the Re-FeFET operates in the phototransistor mode when it is set to the unipolar homojunction state.

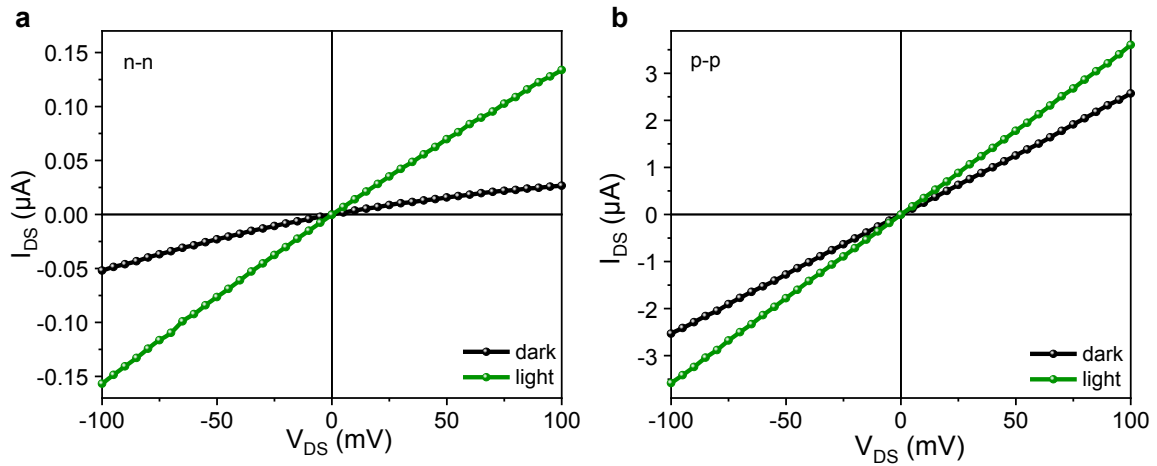

*Figure S4. $I_{DS}$ vs $V_{DS}$ characteristics in dark (black) and under illuminated (green) conditions for **a**, n-n unipolar state, and **b**, p-p unipolar state.*

## Section 5. Ferroelectric logic gate measurements

A ferroelectric logic circuit was built as described in **Figure 5b** and **Figure 5c** of the main manuscript. The K2182A Nanovoltmeter measured the voltage drop across a resistance connected in series with the device. The output voltage $V_{Out}$ across the device is shown in **Figure S5** for two distinct FeFET configurations. **Figure S5.a** (resp. **Figure S5.b**) shows the behavior of the ferroelectric logic when the FeFET is configured in the p-FeFET (resp. n-FeFET) state while storing the downward (resp. upward) polarized state in the Program Gate while applying pulse voltage $V_{PG}$ = -12 V (resp. +12 V). The voltage pulse sequence is applied to the Control Gate to provide the ferroelectric logic hysteresis characteristics. The Re-FeFET logic can switch from Fe-NAND (**Figure S5.a**) to Fe-AND (**Figure S5.b**) operating mode based on the state encoded in the Program Gate.

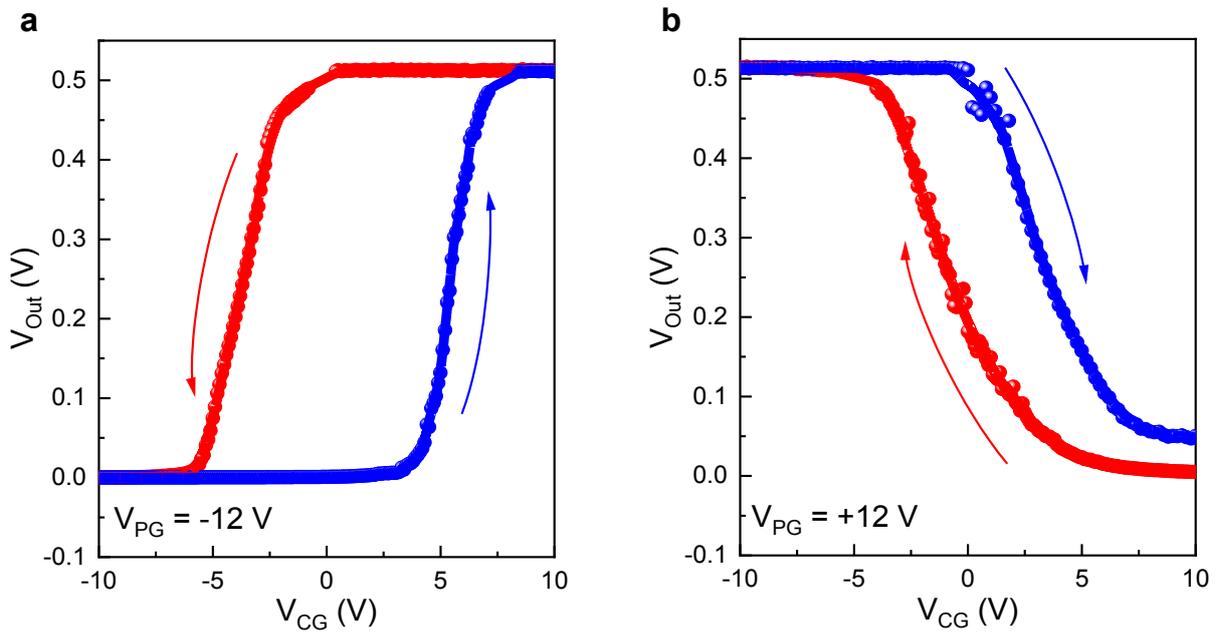

*Figure S5. The variation of $V_{Out}$ with control gate voltage ($V_{CG}$) for two configurations of the Re-FeFET logic circuit. **a,** The downward polarization is stored in the Program Gate while pulsing -12 V. This results in the Fe-NAND $V_{Out}(V_{CG})$ characteristic. **b,** The upward polarization is stored in the Program Gate while pulsing +12 V. This results in the Fe-AND $V_{Out}(V_{CG})$ characteristics.*

## Section 6. Circuit design exploration with Re-FeFETs

We assess the potential of ferroelectric reconfigurable logic circuits based on Re-FeFETs through a circuit design exploration and modeling study. In order to set a framework and terminology for the analysis, we abstract the single device to a set of parameters $\{\{t, PG\}, \{X, CG\}\}$ and set the following axioms:

1. The parameter $t$ represents the non-volatile polarity of the ferroelectric part controlled by the Program Gate $PG$. If $t = 0$, the Re-FeFET behaves as a p-type transistor, while if $t = 1$, it behaves as a n-type transistor. To change the value of $t$, a strong voltage pulse can be applied to the Program Gate. As indicated in the main manuscript section 3, applying a negative voltage pulse to $PG$ leads to the p-FeFET mode ($t = 0$), while applying a positive voltage pulse leads to the n-FeFET mode ($t = 1$).
2. Considering each transistor type (p or n), applying a constant voltage to "$PG$" will shift the regular threshold voltage of the transistor. This makes it possible to create low threshold voltage ("Low $V_{th}$" or LVT) n-type and p-type transistors as well as high threshold voltage ("High $V_{th}$" or HVT) n-type and p-type transistors. Here, $PG = 0$ will be considered to represent the HVT state and $PG = 1$ the LVT state.
3. $X$ corresponds to the non-volatile polarization of the ferroelectric part linked to the Control Gate $CG$. Applying a sufficiently strong negative voltage pulse changes the polarization of the ferroelectric part linked to $CG$ and increases the threshold voltage - this will be noted $X = 0$. The polarization resulting from a positive pulse is noted $X = 1$ and corresponds to a decreased threshold voltage.
4. $CG$ represents the Control Gate of the Re-FeFET. If not used to change the non-volatile polarization $X$, $CG$ will be used as a classical transistor gate in the design exploration.

According to the axioms described above, it is possible to know when the Re-FeFET will conduct ($On$ state) and when it will not ($Off$ state) for each parameter set. For both p and n-mode of the Re-FeFET, three different conduction behaviors can be achieved for regular logic operations within the voltage range used to represent logic '0' and logic '1': $always\ Off/regular/always\ On$. Depending on $X$, the following FeFET behaviors are possible:

- HVT p-mode: $always\ On\ /\ regular$
- LVT p-mode: $regular\ /\ always\ off$
- HVT n-mode: $always\ Off\ /\ regular$
- LVT n-mode: $regular\ /\ always\ On$

This is summarized in Table S1, showing the conduction state of the Re-FeFET for each value of each parameter.

*Table S1: Conduction table of the Re-FeFET according to the value of each parameter $\{\{t, PG\}, \{X, CG\}\}$*

|   |    | $t = 0$ |         | $t = 1$ |         |
|---|----|---------|---------|---------|---------|
| $X$ | $CG$ | $PG = 0$ | $PG = 1$ | $PG = 0$ | $PG = 1$ |
| 0 | 0  | $On$    | $On$    | $Off$   | $Off$   |
| 0 | 1  | $On$    | $Off$   | $Off$   | $On$    |
| 1 | 0  | $On$    | $Off$   | $Off$   | $On$    |
| 1 | 1  | $Off$   | $Off$   | $On$    | $On$    |

To implement non-volatile logic operations using two series-connected Re-FeFETs (presented in the main manuscript **Figure 6c)**, it is possible to search for solutions inside Table S1 with a set of constraints. Each solution represents the parameter set applied to each Re-FeFET of the serial circuit ($Re\text{-}FeFET_1 = \{\{t_1, PG_1\}\{X_1, CG_1\}\}$ and $Re\text{-}FeFET_2 = \{\{t_2, PG_2\}\{X_2, CG_2\}\}$). 6 logic gates are needed to synthetize any logical operation: $NAND, NOR, AND, OR, NOT\ and\ XOR$. This can be reduced to only 3 considering that $AND, OR\ and\ NOT$ can be achieved using $NAND$ and $NOR$ gates. We now demonstrate that the series-connected structure can be used to implement all these logical operations between two operands noted $A$ and B.

As non-volatile logic gates will store one operand inside the ferroelectric layer of the Re-FeFET, it is possible to store either $B$, or $\bar{B}$ ($NOT(B)$). However, to create the most compact logic gates possible, the second operand, which is volatile, is limited to $A$ (or to $\bar{A}$) for the entire circuit. The most difficult logic gate to implement is the $XOR$ because the conduction Table S1 gives directly $NAND$ and $NOR$ truth table (considering $n_1 = V_{dd}$ and $n_2 = gnd$):

- NAND: $Re\text{-}FEFET_1 = \{\{0,0\}, \{B, A\}\}$ and $Re\text{-}FEFET_2 = \{\{1,0\}, \{B, A\}\}$
- NOR: $Re\text{-}FEFET_1 = \{\{0,1\}, \{B, A\}\}$ and $Re\text{-}FEFET_2 = \{\{1,1\}, \{B, A\}\}$

To implement the $XOR$ ($A \oplus B$), we examine the constraints this operation implies on the series-connected structure.

*Table S2: 2-input XOR truth table*

| B | A | $A \oplus B$ |
|---|---|---|
| 0 | 0 | 0 |
| 0 | 1 | 1 |
| 1 | 0 | 1 |
| 1 | 1 | 0 |

From the XOR truth table described in Table S2, we observe the following:
- $A = B = 1, out = 0 \Rightarrow n_1 \text{or} n_2 = 0$. To make sure to only use $A$ (or $\bar{A}$) in the circuit, this implies that $n_1 \text{or} n_2$ must be connected to ground (constant 0). Note that if the circuit uses $\bar{A}$ instead of $A$, this conclusion is derived from the first line of the truth table.
- $A = 0, B = 1, out = 1 \Rightarrow n_1 \text{or} n_2 = 1$. To achieve a logic '1' at the output when $A = 0$, this implies that A should not be connected to $n_1 \text{or} n_2$ because it is necessary to ensure $out = 1$.

Based on this, the following can be considered: $n_1 = V_{dd}$ (logical 1) and $n_2 = Gnd$ (logical 0). With this, it is possible to deduce the conduction state of both Re-FeFETs for each line of the truth table:

$$Re\text{-}FEFET_1 = \{t_1, PG_1, X_1, CG_1\} \Rightarrow \text{Off, On, On, Off}$$

$$Re\text{-}FEFET_2 = \{t_2, PG_2, X_2, CG_2\} \Rightarrow \text{On, Off, Off, On}$$

By crossing these constraints with Table S1, multiple configurations can be found to create a non-volatile $XOR$, for example:

- $Re\text{-}FEFET_1 \in \{\{\{\bar{B}, 0\}, \{1, A\}\}, \{\{\bar{B}, A\}, \{0, 1\}\}, \{\{\bar{B}, A\}, \{1, 0\}\}\}$
- $Re\text{-}FEFET_2 \in \{\{\{B, 0\}, \{1, A\}\}, \{\{B, A\}, \{0, 1\}\}, \{\{B, A\}, \{1, 0\}\}\}$

Two remarks from these solutions are necessary. Firstly, all of them only use $A$ as the volatile operand and other solutions using only $\bar{A}$ also exist. However, this means that no additional inverter is necessary to physically implement this circuit.

In order to simulate the proposed circuit architecture, an equivalent circuit is implemented for a single Re-FeFET device. To reproduce the different behavior of the Re-FeFET described by **Table S1**, a circuit is created under the Cadence environment:

- As ambipolar transistor compact models are not generally available, we proceed by building a two transistor equivalent circuit macromodel: one $PMOS$ and one $NMOS$.
- $PG$ and $X$ induce a threshold voltage shift of the Re-FeFET. Consequently, these parameters can be modeled by a voltage source serially connected to the gate of both transistors. This leads to the three requisite operating modes for each transistor: $always\ Off/\ regular/\ always\ On$. A negative voltage applied to $V_X$ (or $V_{PG}$), will increase the threshold voltage as observed at $CG$, while a positive voltage applied to $V_X$ (or $V_{PG}$) will decrease the threshold voltage as observed at $CG$.
- $CG$ represents the voltage applied to the external gate of the Re-FeFET.
- Finally, transmission gates controlled by $t$ for the $NMOS$ and $\bar{t}$ for the $PMOS$ connected the relevant device drain terminal to the output of the circuit.

All this is summarized in **Figure S6**.

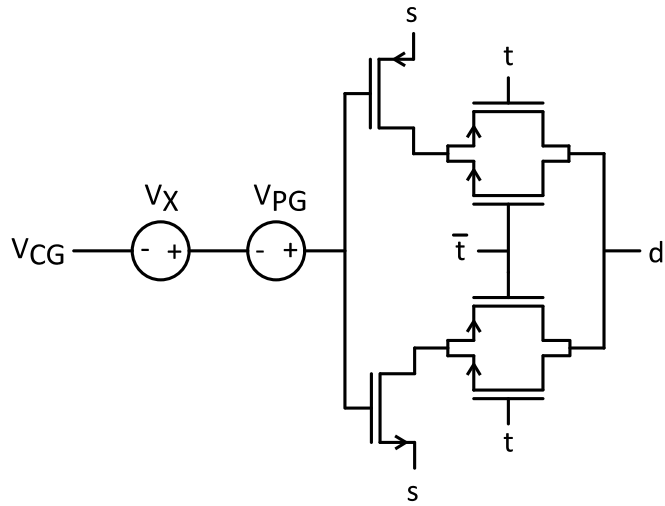

*Figure S6. Re-FeFET emulation using a behavioral equivalent circuit.*

For behavioral simulations, the value applied to $V_X$ and $V_{PG}$ are calibrated to ensure that each transistor (n and p) can achieve the 3 conduction modes ($always\ off$, $regular$, and $always\ on$) within the voltage range used to represent logic '0' and logic '1'. As the supply voltage for the technology used for simulation is 1.2 V, and as this voltage also corresponds to logic '1', we choose a total threshold voltage shift $V_{sh}$ of $-1.2V$, $0V$ or $1.2V$ as observed from the external gate of the component. $V_{sh}$, $V_X$, and $V_{PG}$ are linked by the following expression:

$$V_{sh} = V_X + V_{PG}$$

$V_X$ (resp. $V_{PG}$) can only take two values, one representing $X = 0$ (resp. $PG = 0$) and one representing $X = 1$ (resp. $PG = 1$). Consequently, we obtain:

$$\begin{cases} V_X \in \{-0.6V, 0.6V\} \\ V_{PG} \in \{-0.6V, 0.6V\} \end{cases} \Rightarrow V_{sh} \in \{-1.2V, 0V, 1.2V\}$$

With this, we achieve for the p-type transistor (symmetrical behavior is achieved for the n-type transistor):

- Always ON: $\begin{cases} X = 0 \Rightarrow V_X = -0.6V \\ PG = 0 \Rightarrow V_{PG} = -0.6V \end{cases} \Rightarrow V_{sh} = -1.2V$
- Regular: $\begin{cases} X = 0 \Rightarrow V_X = -0.6V \\ PG = 1 \Rightarrow V_{PG} = 0.6V \end{cases} \Rightarrow V_{sh} = 0V$ or $\begin{cases} X = 1 \Rightarrow V_X = 0.6V \\ PG = 0 \Rightarrow V_{PG} = -0.6V \end{cases} \Rightarrow V_{sh} = 0V$
- Always OFF: $\begin{cases} X = 1 \Rightarrow V_X = 0.6V \\ PG = 1 \Rightarrow V_{PG} = 0.6V \end{cases} \Rightarrow V_{sh} = 1.2V$